# First Principles Study of 2D Ring-Te and its Electrical Contact with Topological Dirac Semimetal

Jaspreet Singh and Ashok Kumar[*]

*Department of Physics, Central University of Punjab, VPO Ghudda, Bathinda, 151401, India*

(February 03, 2023)

*Email: ashokphy@cup.edu.in

**Abstract**

In recent years, researchers have manifested their interest in the two-dimensional (2D) mono-elemental materials of group-VI elements because of their excellent optoelectronic, photovoltaic and thermoelectric properties. Despite the intensive recent research efforts, there is still a possibility of novel 2D allotropes of these elements due to their multivalency nature. Here, we have predicted a novel 2D allotrope of tellurium (ring-Te) using density functional theory. Its stability is confirmed by phonon and ab-initio molecular dynamics simulations. The ring-Te has an indirect band gap of 0.69 eV (1.16 eV) at PBE (HSE06) level of theories and undergoes an indirect-direct band gap transition under the tensile strain. The higher carrier mobility of holes ($\sim 10^3 cm^2 V^{-1} s^{-1}$), good UV-visible light absorption ability and low exciton binding (~0.35 eV) of ring-Te gives rise to its potential applications in optoelectronic devices. Further, the electrical contact of ring-Te with topological Dirac semimetal (sq-Te) under the influence of electric field shows that the Schottky barriers and contact types can undergo transition from p-type to n-type Schottky contact and then to ohmic contact at higher electric field. Our study provides an insight into the physics of designing high-performance electrical coupled devices composed of 2D semiconductors and topological semimetals.

# Introduction

The ever-expanding family of 2D materials has revolutionized the material science field with their exotic properties. 2D materials such as graphene,[1-3] boron nitride,[4] transition metal dichalcogenides (TMDs),[5,6] borophene,[7] phosphorene[8,9] and germanene based 2D materials[10,11] have established a key place in the fields of optoelectronics,[6,12] photonics,[13] and energy storage.[14] Graphene, the first 2D monolayer, has intriguing properties such as high mechanical strength[2] and ultrahigh carrier mobility,[15] making it a superior material in futuristic nanodevice applications. However, the lack of band gap at the Dirac cone in graphene restricts its applications in the electronic industry, especially in field-effect transistors (FETs). With the advancements in research methods, scientists have introduced many ways such as functionalization,[16,17] defects,[18,19] strain,[20] and electric-field[21] engineering to open the band gap of graphene and utilize it as a semiconducting material in device applications.

However, in the past few years, graphene based 2D van der Waals heterostructures (vdWHs) such as graphene/phosphorene,[22] graphene/WS$_2$,[23] graphene/Te,[24] graphene/MoSi$_2$N$_4$,[25] graphene/InP$_3$[26] and graphyne/graphene[27] are employed as FETs in which the graphene acts as a metal electrode to contact the 2D semiconducting channel. The charge transport in these devices depends upon the Schottky barrier formed at the interface of vdWH. The Schottky barrier can be tuned with the external electric field and mechanical strain to achieve a desirable ohmic contact at the interface to reduce the power dissipation.[28,29] In addition to graphene based vdWHs, other vdWHs such as NbS$_2$/MoSi$_2$N$_4$,[25] ZT-MoSe$_2$/PdSe$_2$,[30] ZT-MoSe$_2$/Phosphorene,[31] planar/buckled-silicene[32] etc. are also studied extensively in device applications.

Recently, adding to the field of vdWHs, Cao *et al*. proposed a new method of employing the topological Dirac semimetal as an electrical contact to 2D semiconductors.[33] They have constructed a highly efficient electrical contact between bilayer Na$_3$Bi and MoS$_2$/WS$_2$ monolayers with shallow Schottky barrier heights. For high-performance 2D FETs, the semiconducting channel material should also have high carrier mobility, mechanical strength and tunable electronic properties. In the past few years, 2D mono-elemental materials of group VI-b (Tellurium (Te), Selenium (Se), Sulfur (S)) elements have emerged as promising materials overcoming the shortcomings such as lack of band gap, low carrier mobilities, air stability of existing 2D materials. These 2D materials are predicted to have various allotropic forms such as α-Te (Se), β-Te, γ-Te and sq-Te (Se).[34-36] Besides these pristine materials, the Janus materials such as α-Se$_2$Te, α-Te$_2$Se, α-Te$_2$S[37-39] are also predicted based on first principles theory. Owing to the multivalency nature of these elements, there is still a greater possibility for the existence of their novel allotropes that might have superior properties.

In this work, we have reported the novel 2D ring structure of Te referred as ring-Te. The stability of ring-Te is confirmed by the calculations of phonons, ab-initio molecular dynamics (AIMD) and mechanical properties. For device applications, we have constructed the vdWH of ring-Te with sq-Te (sq-Te/ring-Te). The semi-metallic properties of sq-Te and its possibility of being grown on a proper substrate[35] make it a promising material to be used as an electrical contact to 2D semiconductors in vdWHs. Further, we have investigated the interfacial electronic properties of sq-Te/ring-Te and modulated its Schottky barriers and contact types with the vertical electric field.

## Computational Details

We have used the density functional theory (DFT) as implemented in the Quantum Espresso package[40] to study the structural and electronic properties of monolayer ring-Te and its vdWH with sq-Te. The projector-augmented wave potentials (PAW)[41] and generalized gradient approximations proposed by Perdew-Burke-Ernzerhof (PBE)[42] are used to take care of electron-ion interactions and exchange-correlation energies, respectively. The energy cutoff of 60 Ry and k-point mesh of 12x12x1 is used for ground state energy calculations. The Heyd–Scuseria–Ernzerhof (HSE06) method[43] is also used to obtain the more accurate band gap of monolayer ring-Te. The force and energy convergence of $10^{-6}$ eV/Å and $10^{-7}$ eV, respectively, are used for the structural relaxations. The spin-orbit coupling (SOC) effects are included in calculating the electronic properties due to the heavy Te atom. The DFT-D3 method proposed by Grimme[44] is used to treat the weak van der Waals interactions in the heterostructure. To study the variation of electronic band structures of heterostructures under the external electric field, we have employed a saw-like potential as implemented in Quantum Espresso package. A vacuum slab of ~12 Å is employed to eliminate the interlayer interactions between neighboring layers. The density functional perturbation theory (DFPT) has been used to obtain the phonon dispersion by taking the 8x8x1 k-point mesh and 4x4x1 q-point mesh with convergence threshold of $10^{-18}$ Ry for phonons self-consistency calculations.[45] We have performed the AIMD simulations controlled by a Nose-Hoover thermostat[46, 47] for 5000 fs with a time step of 1 fs. YAMBO code[48] interfaced with the Quantum Espresso package is used to evaluate the optical properties with the $G_0W_0$ and Bethe-Salpeter equation (BSE) method. A plasmon-pole model is adopted for the dynamic screened interactions and cut-off energies of 60 Ry and 6 Ry are used for the exchange and correlation part of the self-energy, respectively. The 16x16x1 k-point mesh and a total of 200 bands are taken for the calculations of optical spectra. For the absorption spectra, we have taken a total of 4 valence bands and 3 conduction bands to solve the Bethe–Salpeter equation.

## Results and Discussion

**Structure and Stability**

The relaxed 2D structure of ring-Te is shown in Fig. 1(a), which is similar to PCF-graphene[49] from top. The repetitive square unit cell constitutes 12 atoms with a relaxed lattice constant of 8.30Å. Out of the 12 atoms, there are two types of Te atoms marked by position 1 ($Te_1$) and 2 ($Te_2$) in the side view (Fig. 1(a)) with coordination numbers 4 and 3, respectively. The bond length between $Te_1$ and neighboring Te atoms ($Te_1$-Te) is 3.02 Å, while the bond length between $Te_2$ atoms ($Te_2$-$Te_2$) is 2.80 Å. Consequently, the agreement of asymmetric bond lengths with the literature[34] reveals the metal-ligand type bonding in ring-Te. Note that the buckling height '$t_1$' of monolayer is calculated to be 4.22 Å.

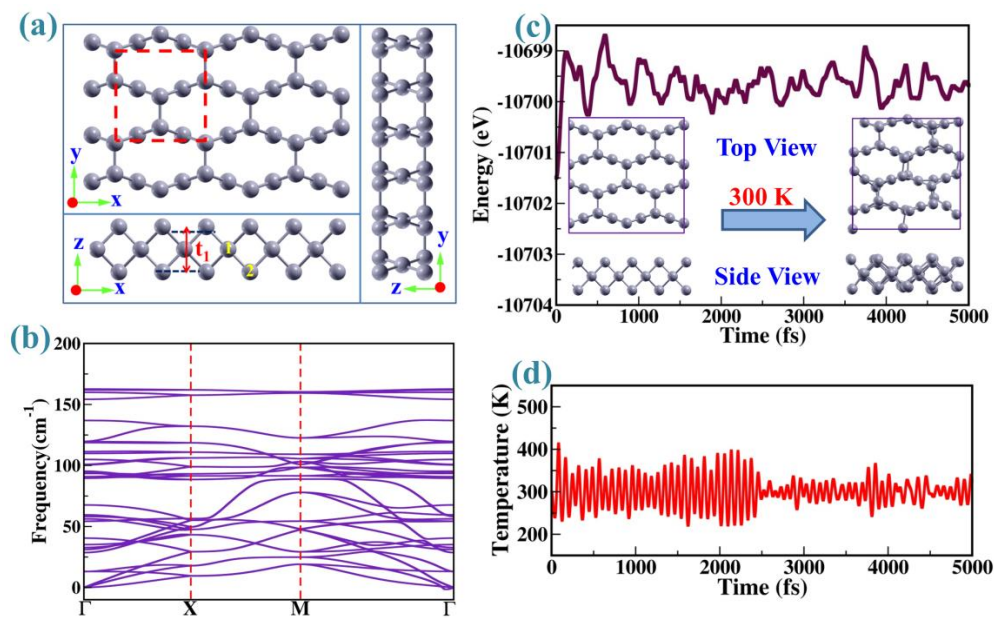

**Fig. 1** Ring-Te: **(a)** Different views of crystal structure, **(b)** phonon spectrum, variation of **(c)** energy and **(d)** temperature with time steps. The monolayer structure at the end of 5000 fs is shown in the inset of **(c)**.

Now we examine the energetic stability of ring-Te by calculating the cohesive energy which comes out to be 3.04 eV/atom that is comparable with the most stable phase of Te i.e., α-Te (3.05 eV/atom),[50] revealing relative energetical stability of ring-Te. Following that, the dynamical stability analysis with phonon spectrum calculations suggests no imaginary frequencies in phonon band dispersion over the irreducible Brillouin zone of ring-Te (Fig. 1(b)). The AIMD simulations reveals that the energy and temperature w.r.t time steps fluctuate around the steady level (Fig. 1(c, d)) indicating its thermal stability at 300K. The negligible distortion in the final structures (inset of Fig. 1(c)) after the 5000 fs confirms the stability of ring-Te in an isolated form.

**Mechanical Properties**

Further, we analyze the mechanical properties of ring-Te by investigating its response under the uniaxial strain ($\varepsilon = \frac{l - l_0}{l_0}$, where $l$ and $l_0$ are the lattice constants for strained and relaxed systems). The axis orthogonal to the stressed axis is entirely relaxed to ensure the minimum stress in the orthogonal direction and the applied strain is uniaxial. The maximum stress that a given structure can bear without breaking of atomic bonds is known as the ideal tensile strength (ITS). The ITS is calculated from the strain-stress relation using the ab-initio pseudopotential density functional method which is originally designed for three-dimensional crystals.[51, 52] For 2D crystals, the calculated ITS is rescaled with the parameter "$z/t_0$" ($z$ is the vacuum length and $t_0$ is the effective thickness of the relaxed system) to avoid the effect of vacuum in the z-direction and make the results directly comparable with experiments.[53, 54] The calculated ITS is 1.38 GPa and 2.32 GPa, corresponding to 12% and 26% critical strains along x- and y-directions, respectively (Fig. 2(a)) after which the ring-Te enters into the plastic region. The values of ITS are less than that of other 2D Te allotropes i.e. α-Te (7.12 GPa), β-Te (11.43 GPa) and γ-Te (4.49 GPa).[55] However, its critical strain values are comparable to that of γ-Te. The values of Young's modulus are calculated to be 30.12 GPa and 11.27 GPa along x- and y-direction, respectively, which are lower than that of α-Te, β-Te, and γ-Te.[55] The lower values of Young's modulus with higher critical strain limits along the y-direction indicate the mechanical flexibility of ring-Te. The variation of transverse strain with axial strain is depicted in Fig. 2(b), from which we have calculated the Poisson ratio values of 0.34 and 0.20 along x- and y-directions, respectively. The above discussion reveals the anisotropic mechanical properties of ring-Te having a superior flexibility in y-direction as compared to x-direction.

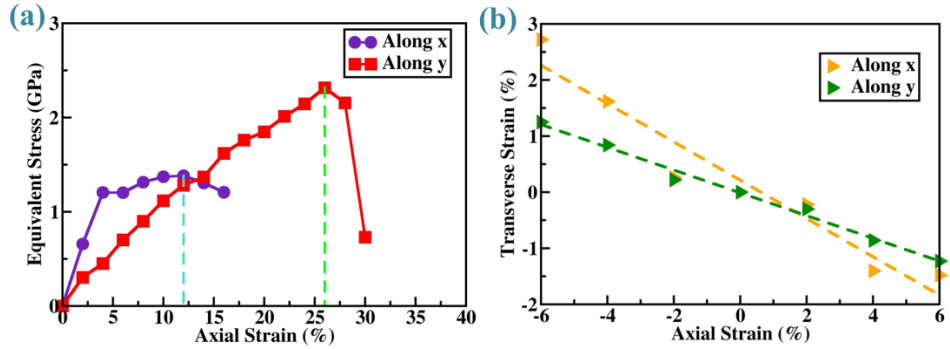

**Fig. 2** Ring-Te: Variation of **(a)** equivalent stress and **(b)** transverse strain with applied axial strain, respectively.

To get deeper insight into the anisotropic nature of ring-Te, we have further analyzed bonding characteristics in terms of bond angles and bond lengths (Fig. 3). The response of ring-Te under the tensile strain along the x-direction reveals the opposite change in the bond angles ($\Theta_1$, $\Theta_2$) and bond lengths ($L_1$, $L_3$). After the slight decline in '$\Theta_1$' up to 4% strain, there is a rapid increment in '$\Theta_1$' upto ~ 8% before it becomes nearly constant (Fig. 3(a)). In contrast, the continuous declination in '$\Theta_2$' values is observed with a maximum percentage change of 19.19 obtained at a maximum strain value of 12%. Also, a contraction and extension in the bond length '$L_1$' and '$L_3$', respectively, are observed. These opposite changes cause the small distortions (more prominent after the 4% strain) in ring-Te due to the opposite movement of atoms in different planes shown in Fig. S1(a), ESI. This distortion is the reason for declination in the stress-strain curve slope observed after the 4% tensile strain along the x-direction (Fig. 2(a)). The bond angles 'Θ$_3$' and 'Θ$_5$' have shown the expected rise while the bond angle 'Θ$_4$' and bond length 'L$_2$' remains almost constant as these are perpendicular to applied strain (Fig. 3(a, b)).

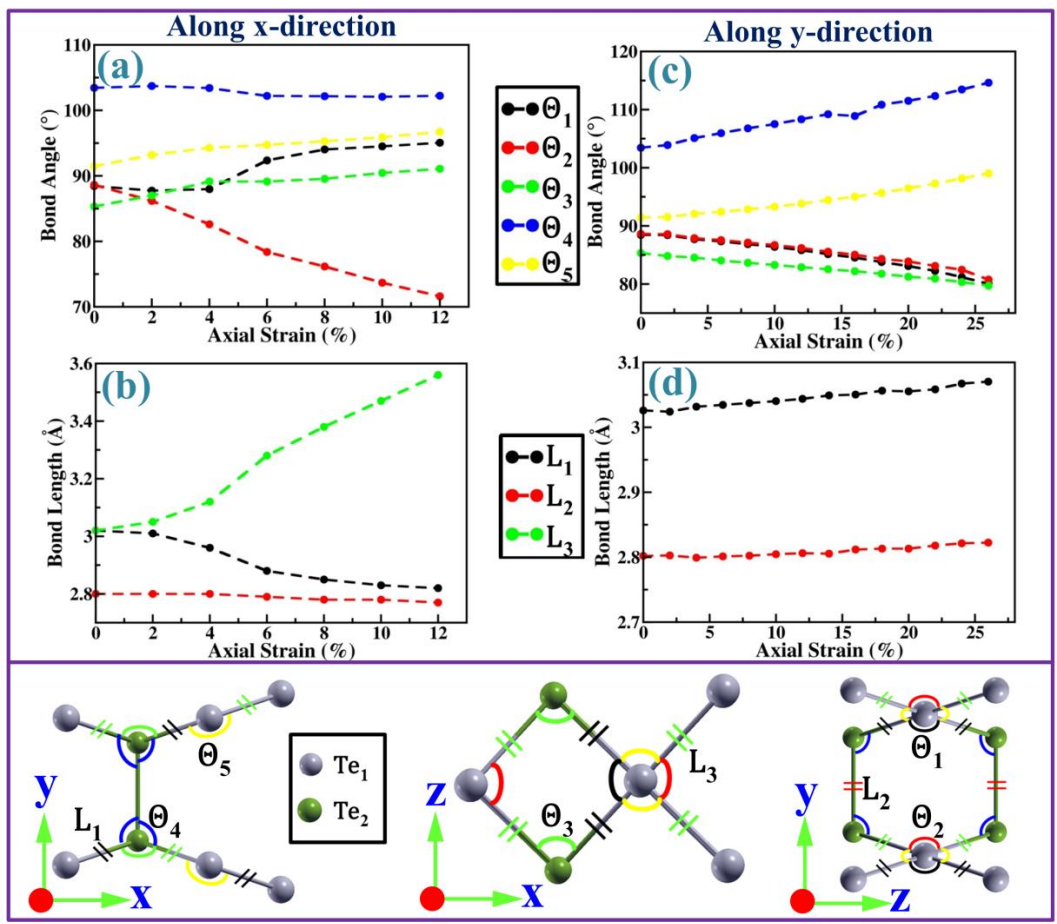

**Fig. 3** Variation of **(a, c)** bond angles (Θ$_1$, Θ$_2$, Θ$_3$, Θ$_4$, Θ$_5$) and **(b, d)** bond lengths (L$_1$, L$_2$, L$_3$) of ring-Te with uniaxial applied strain. The bond length L$_1$ and L$_3$ along y-direction show similar variation with strain, therefore, only L$_1$ variation with strain along y-direction is shown.

When the strain along the y-direction is applied, the bond lengths showed minimal change (less than 2%) (Fig. 3(d)). All the bond angles have exhibited the expected changes, with the most prominent rise of 10.81% in bond angle 'Θ$_4$' (Fig. 3(c)). Thus, the tensile strain along the y-direction has only flattened the buckling height by 5.45% without the extension of Te$_1$-Te and Te$_2$-Te$_2$ bonds shown in Fig. S1(b), ESI. It leads to the linear increase in the equivalent stress with tensile strain up to 26% along the y-direction (Fig. 2(a)). The bond angles and lengths at maximum tensile strain along x- and y-directions, respectively, with their percentage changes, are listed in Table S1, ESI.

From the above discussion, it is clear that when the ring-Te is strained along x-direction, it causes the abrupt changes in the bond angles and bond lengths as we can see from the behavior of $\Theta_1$, $\Theta_2$, $L_1$ and $L_3$. In contrast to it, the gradual changes are observed in bond lengths and bond angles when the ring-Te is strained along y-direction. This implies that the small strain along the x-direction causes larger distortion in the crystal structure of ring-Te as compared to strain along y-direction that leads to its anisotropic behavior. This anisotropic response is due to the more puckered crystal structure of ring-Te along y-direction (Fig. 1(a)). The total charge density profiles of ring-Te also reveal that there is an overlapped charge distribution along y-direction as compared to the more localized charge distribution along x-direction (Fig. S2, ESI). This type of distribution leads to the stronger bonding-like states along y-direction as compared to the anti-bonding-like states along x-direction, respectively. This causes the stronger mechanical stability of ring-Te along y-direction that leads to its higher ITS and critical strain along y-direction as compared to x-direction.

**Electronic Properties**

Having analyzed the stability of ring-Te, we have now explored its electronic properties. Fig. 4(a) suggests the semiconductor nature of ring Te with an indirect band gap of 0.69 eV having valence band maximum (VBM) at Γ point and conduction band minimum (CBM) at X point. A comparison of its electronic properties without the inclusion of SOC (Fig. S3(a), ESI) suggests that there is not much difference in band gap, however the SOC effect has been taken to maintain the consistency. We have analyzed the orbital contribution around the Fermi level in terms of PDOS (Fig. 4). The SOC effect is due to the interaction between total spin (s) and orbital (l) angular momentum. These angular momentums are combined via l-s coupling give rise to the total angular momentum (j = (l + s), (l + s - 1)…..(l – s)). VBM mainly composed of p-orbitals of both types of Te atoms and the CBM is primarily contributed by p-orbitals of $Te_2$ atoms, with higher contributions from the orbitals corresponding to higher total angular momentum (j = 3/2).

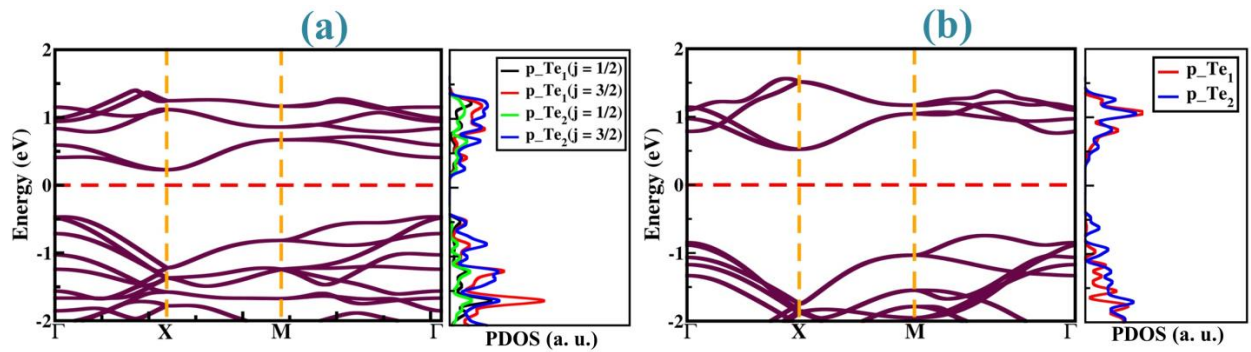

**Fig. 4** Electronic band structure and PDOS of ring-Te **(a)** with PBE functional including SOC effect **(b)** with HSE06 functional. The Fermi level is set at 0.

To overcome the limitation of underestimation of band gap obtained by PBE functional, we have obtained the band gap of 1.16 eV of isolated monolayer with hybrid HSE06 functional with overall dispersion almost remains the same (Fig. 4(b)). Also, the value of band gap calculated at $G_0W_0$ level of theory is 1.34 eV. The $G_0W_0$ electronic band structure of ring-Te is given in Fig. S3(b), ESI.

**Modulation of Electronic Properties under Strain**

Next, we have analyzed the electronic properties of ring-Te under the uniaxial strain. Since the maximum strain limits (critical strains) are calculated to be 12% and 26%, respectively, along x and y-directions, therefore, we have checked the variation of electronic band structures of the ring-Te under the different strain range along x and y-directions. The variation of band gaps with uniaxial strain is shown in Fig. 5(a, b) and corresponding band structures are given in Fig. S4, S5, ESI. Our results suggest that with the increase in tensile strain in both directions, the indirect band gap of the monolayer increases. The initial smaller dip in the band gap with tensile strain along the x-direction is due to the slight distortion in structure after the 4% tensile strain. Note that at the critical strain of 16% (marked in Fig. 5(b)) along the y-direction, the indirect to direct band gap transition occurs with CBM shifting from X to Γ point (Fig. S5, ESI). Note that at 20% strain, we have checked the relative stability of ring-Te, by calculating its cohesive energy value. The cohesive energy value of ring-Te at 20% strain comes out to be 3.01 eV/atom which is not much different than that of the cohesive energy value of 3.04 eV/atom for strain-free ring-Te that indicate the energetical stability of ring Te at higher strains.

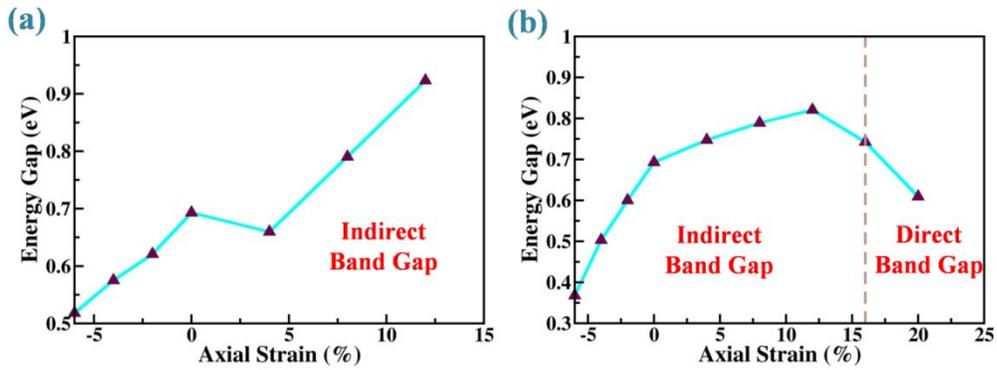

**Fig. 5** Variation of band gap of ring-Te with applied axial strain along **(a)** x- and **(b)** y-direction, respectively.

Further, we have plotted the band decomposed charge densities of VBM (at Γ point) and CBM (at X and Γ point) at different strain values to reveal the reason behind the indirect-direct band gap transition (Fig. S6, ESI). The charge density plots clearly indicates the opposite trend in charge densities of CBM around X and Γ with increase in strain values. At point X, with the increase in strain, the charge gets depleted from the surface Te atoms ($Te_2$) and gets accumulated around the middle atoms ($Te_1$) and vice versa at point Γ. This results in change in the position of CBM from X to Γ point and hence the nature of band gap. Note that the CBM is mainly contributed by p-orbitals of $Te_2$ atoms.

**Carrier Mobility**

For the utilization of materials in electronic devices, the materials should have high carrier mobility. We have calculated the carrier mobility ($\mu$) using the modified formulation of the longitudinal acoustic phonon limited model proposed by Lang et al.[56] as:

$$\mu_{ax} = \frac{e\hbar^3 \left(\frac{5C_x^{2D}+3C_y^{2D}}{8}\right)}{k_B T m_{ax}^{\frac{3}{2}} m_{ay}^{\frac{1}{2}} \left(\frac{9E_{ax}^2 + 7E_{ax}E_{ay} + 4E_{ay}^2}{20}\right)} \quad (1)$$

where, $C^{2D}$, $m$ and $E$ is the modulus of elasticity, effective mass and deformation potential of carriers ($a$= electron or hole). With this modified formulation, mobility anisotropy has been more accurately measured, which depends mainly on anisotropy in the effective mass rather than the anisotropy in deformation potential and elastic modulus.[57, 58] These parameters have been evaluated by the fitting curves between energy and strain, as shown in Fig. S7(a, b, c), ESI. The smaller effective masses of holes result in higher carrier mobility of the holes than electrons by order of ~10. The anisotropic ratio ($R_{ani} = \frac{Max(\mu_x, \mu_y)}{Min(\mu_x, \mu_y)}$) of 1.48 and 1.97 for holes and electrons indicates the anisotropy in their carrier mobilities. This anisotropy arises due to the small contributions from the anisotropy in elastic modulus and deformation potentials. The carrier mobility of holes ($10^3 cm^2 V^{-1} s^{-1}$) is comparable to α-Te and β-Te,[34] while the carrier mobility of electrons is of the order of $10^2 cm^2 V^{-1} s^{-1}$ (Table1).

**TABLE 1** Modulus of Elasticity ($C^{2D}$), Effective Mass ($m$), Deformation Potential ($E^d$) and Carrier Mobility ($\mu$) of ring-Te at 300K.

| Direction | $C^{2D}$ (Jm$^{-2}$) | $m$ ($m_0$) | $E^d$ (eV) | $\mu$ ($10^2 cm^2 V^{-1} s^{-1}$) |
|---|---|---|---|---|
| X | 23.68 | $m_e$= 1.95<br>$m_h$= 0.24 | $E_d^e$= 1.622<br>$E_d^h$= 3.242 | $\mu_e$= 0.95<br>$\mu_h$= 12.59 |
| Y | 14.58 | $m_e$= 1.95<br>$m_h$= 0.24 | $E_d^e$= 0.026<br>$E_d^h$= 0.954 | $\mu_e$= 1.88<br>$\mu_h$= 18.64 |

**Optical Properties**

Considering the semiconducting nature of ring-Te, we have evaluated its optical properties for applications in optoelectronic devices. We have employed the $G_0W_0$+BSE method[59] to obtain the imaginary part ($\varepsilon_2$) of dielectric constant from which we have calculated the optical absorbance ($A(\omega)$)[60] as:

$$A(\omega) = \frac{\omega}{c} L \varepsilon_2(\omega) \quad (2)$$

where L is the cell length in the z-direction.

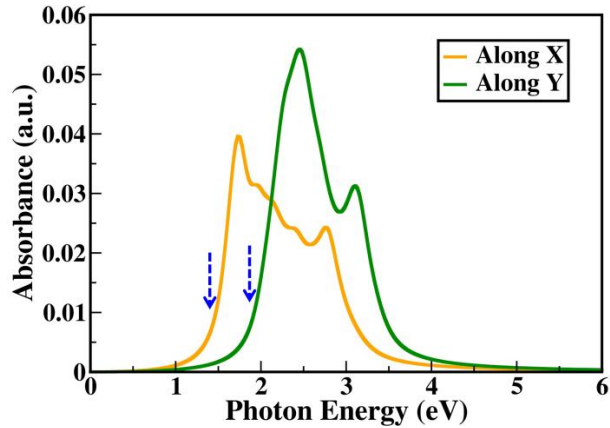

**Fig. 6** Optical absorbance of ring-Te using the $G_0W_0$+BSE method. Vertical arrows show the exciton absorption peaks.

Fig. 6 shows the anisotropic absorbance spectra covering the UV-visible region (~1.5-4 eV). The first exciton peak (along x-direction) lies in the visible region at around 1.40 eV represents the bound exciton state with a binding energy of 0.35 eV. Note that the exciton binding energy is obtained by the difference between the minimum direct quasiparticle band gap and energy of first excitonic peak (optical gap). In addition to that, the exciton first peak along the y-direction also lies in the visible region i.e., around the 1.85 eV. The obtained exciton binding energy is lower than that of α-Te (0.37 eV), α-Se (0.53 eV), β-Te (0.84 eV)[61, 62] that leads to the easier separation of charge carriers which implies its applications in light-harvesting devices. Thus, the semiconducting nature, high carrier mobility and UV-visible light absorption in the ring-Te reveals its potential applications in various fields such as nano-electronic devices, solar conversion applications, etc.

**Electrical Contact between ring-Te and sq-Te**

**Structural and Electronic Properties of vdWH (sq-Te/ring-Te)**

Now we have constructed the vdWH of ring-Te with sq-Te for the investigation in the device applications. The sq-Te has a square unit cell with optimized lattice constant and thickness 't$_2$' of 4.11 Å and 0.90 Å, respectively (Fig. 7(a)). These lattice parameters are in accordance with previously reported values (4.08 Å and 0.92 Å),[35] (4.18 Å and 0.88 Å).[63] The electronic band structure shows a semi-metallic nature having an anisotropic Dirac cone along Γ-M$_1$ direction that gets open by the magnitude of ~0.12 eV (0.16 eV[35]) with the inclusion of SOC effects (Fig. 7(b), Fig. S3(c), ESI). The sq-Te has been predicted to have the non-trivial topological properties [35] that we have utilized to form the electrical contact with ring-Te. We have constructed the heterostructure with a lattice mismatch of 0.96 % by stacking the 1x1 unit cell of ring-Te and 2x2 unit cells of sq-Te along the z-direction to minimize the effect of mechanical stress on calculated results (Fig. 7(c)). The calculated interlayer distance of 3.31 Å is larger than the sum of covalent radii of Te atoms, implying the weak vdW interactions between the layers. The calculated negative binding energy ($E_b$= ($E_{hetero}$ −$E_{ring\text{-}Te}$− $E_{sq\text{-}Te}$)/N, where $E$ and n, represent the respective energies and the total number of atoms, respectively.) of 0.28 eV/atom hints at the energetical stability and feasibility of experimental fabrication of heterostructure. The intrinsic electronic properties of ring-Te and sq-Te are well preserved in the projected heterostructure band structure, which is an important aspect of vdWHs (Fig. 7(d)). However, due to the more pronounced splitting of the valence band of sq-Te at the Dirac cone (~0.07 eV), the VBM crosses the Fermi level. In addition to this, an upward shift of band structure of ring-Te is observed due to the small interactions between the constituent's monolayers. Note that the band gap of ring-Te decreases to 0.66 eV.

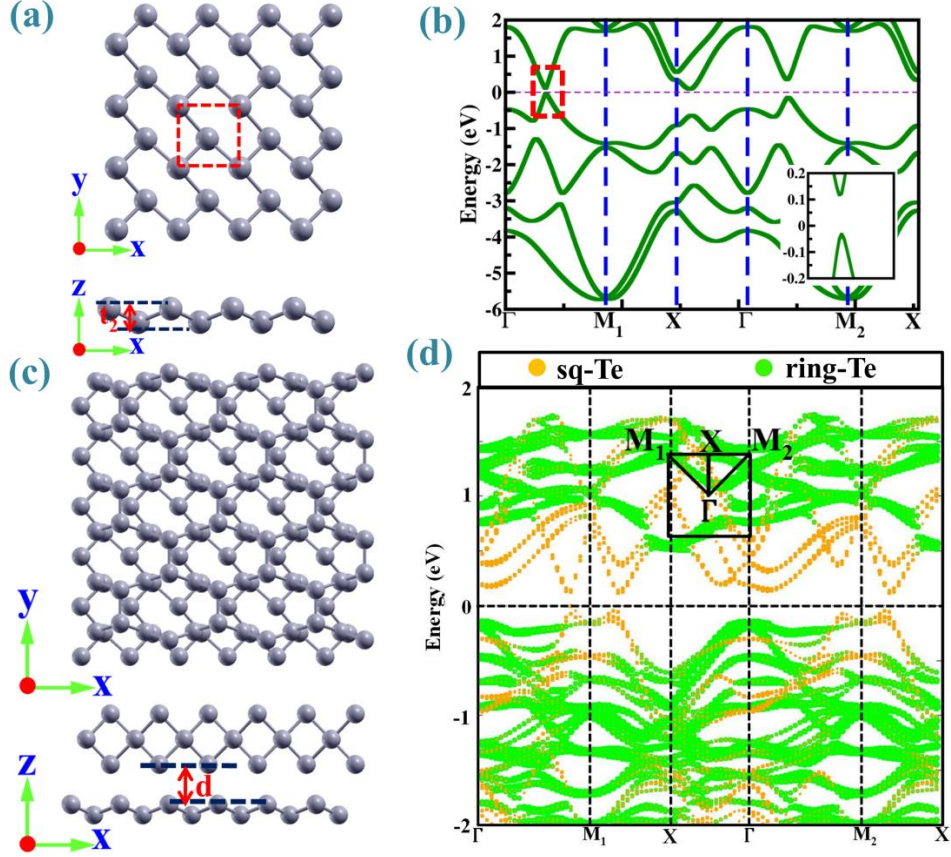

**Fig. 7** Atomic and Electronic band structure with PBE+SOC of sq-Te **(a, b)** and heterostructure (sq-Te/ring-Te) **(c, d)**, respectively. Fermi level is set at 0 eV.

Next we have studied the transportation of charge carriers in FET based on our proposed heterojunction (sq-Te/ring-Te). Fig. 8(a) shows the schematic of FET in which the sq-Te acts as a source and the ring-Te acts as a channel. In the transportation, when the charge carriers move from the stacked region of ring-Te and sq-Te to the region of free-standing ring-Te (channel), there will be small band bending ($\Delta E_f$). This small band bending is obtained by the difference between the work functions of heterojunction ($W_H$) and ring-Te ($W_S$) as:

$$\Delta E_f = W_H - W_S \quad (3)$$

Work functions are calculated as the difference between the Fermi energy and vacuum level as shown in Fig. 8(b). The band bending decides the channel properties at equilibrium i.e. if $\Delta E_f >$ 0, the channel is p-type and the charge will transfer from semiconductor (channel) to heterostructure (stacked region) and vice versa if $ΔE_f < 0$, making the channel n-type. In our case, the calculated value of $ΔE_f$ is 0.31 eV (>0), making the holes major charge carriers (shown in Fig. 8(a)) and hence the channel is p-type (Fig. 8(b)). Note that similar behavior is observed in phosphorene/graphene,[22] PdSe$_2$/ZT-MoSe$_2$[30] based devices.

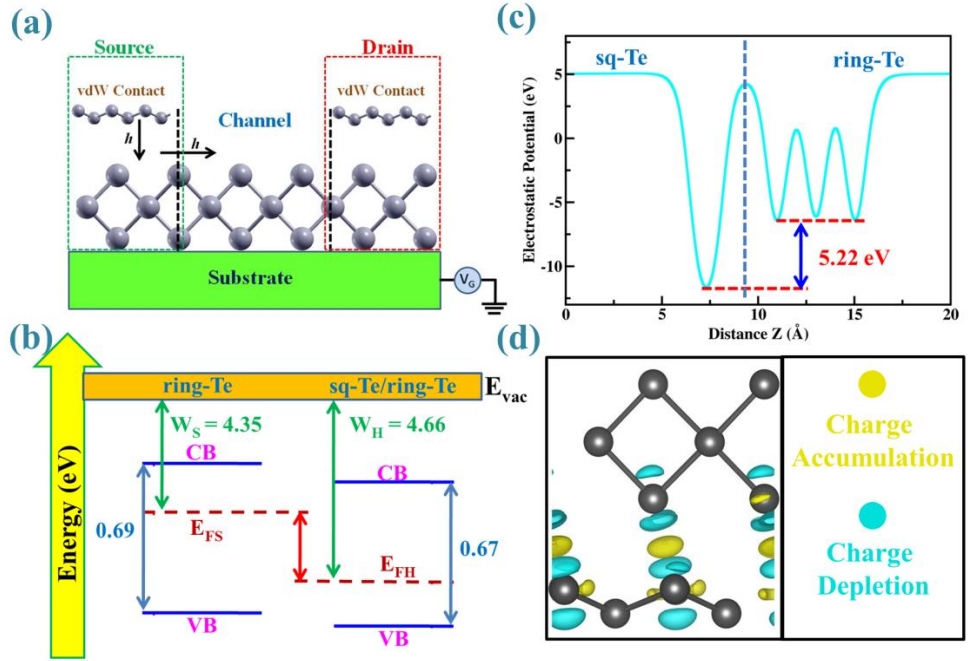

**Fig. 8(a)** The schematic of FET based on vdWH (sq-Te/ring-Te). **(b)** Band alignments of ring-Te and heterostructure (sq-Te/ring-Te). W, E$_F$, E$_{vac}$, CB and VB represent the work function, Fermi level, vacuum level, conduction band and valence band, respectively. **(c)** Electrostatic potential along the z-direction and **(d)** charge density difference of the sq-Te/Ring-Te heterostructure at the equilibrium state, respectively. The isosurface value is 0.0005 eÅ$^{-3}$.

Further, we have determined the Schottky barrier height (SBH) at the heterojunction interface using the Schottky Mott rule.[64, 65] Because of the absence of chemical bonds at the interface of heterojunction, the Fermi level pinning effect is neglected and the Schottky barriers (SBs) are calculated as:

$$\phi_n = E_{CBM} - E_f \qquad (4)$$

$$\phi_p = E_f - E_{VBM} \qquad (5)$$

where $\phi_n$ and $\phi_p$ are the n- and p-type Schottky barriers. Our calculated $\phi_n$ and $\phi_p$ are 0.47 eV and 0.20 eV, respectively, making the heterojunction a p-type Schottky contact at an equilibrium interlayer distance of 3.31 Å. To get more insight into the charge distribution and transfer at the interface in equilibrium, we have calculated the electrostatic potential and charge density difference (CDD) profiles. The potential difference of 5.22 eV between the monolayers in heterojunction implies the induced electrostatic field at the junction that causes the negative charge transfer from ring-Te to sq-Te (Fig. 8(c)). The CDD ($\Delta\rho = \rho_H - \rho_{ring-Te} - \rho_{sq-Te}$, where $\rho_H$, $\rho_{ring-Te}$ and $\rho_{sq-Te}$ are charge densities of the heterostructure, ring-Te and sq-Te, respectively) shows the depletion of charges from the ring-Te, that gets accumulated between the interface region and on sq-Te indicating the p-type Schottky contact (Fig. 8(d)).

**Tuning of Schottky Barriers under Electric Field**

Further, we have tuned the SBs to induce an ohmic contact with the application of an external perpendicular electric field (can be applied through the gate voltage (V$_G$) shown in Fig. 8(a)) as done in the literature.[22, 24, 66] We have applied the vertical electric field that varies from -1.8 V/Å to +1.6 V/Å along the z-direction, whose positive direction has been assumed from the sq-Te to ring-Te. Fig. 9 shows the band edges and band structure variation under the applied electric field. From Fig. 9(a), it is clear that p-type Schottky contact remains intact under the positive electric field because of the decrement of $\phi_p$. Note that minimum $\phi_p$ has been obtained ~0.08 eV at 1.6 V/Å after which the band structure of ring-Te not remain preserved. The opposite trend is noticed as expected when we applied the negative electric field. The negative electric field reduces the $\phi_n$, as a result the transformation from p-type contact to n-type contact happens at -0.8 V/Å. On further increasing the negative electric field, the $\phi_n$ reduces to the negative value (- 0.034 eV) at -1.6 V/Å, hence the n-type Schottky contact changes to n-type ohmic contact. From Fig. 9(a), it is clear that p-type Schottky contact remains intact under the positive electric field because of the decrement of $\phi_p$. Note that minimum $\phi_p$ has been obtained ~0.08 eV at 1.6 V/Å after which the band structure of ring-Te not remain preserved. The opposite trend is noticed as expected when we applied the negative electric field. The negative electric field reduces the $\phi_n$, as a result the transformation from p-type contact to n-type contact happens at -0.8 V/Å. On further increasing the negative electric field, the $\phi_n$ reduces to the negative value (-0.034 eV) at -1.6 V/Å, hence the n-type Schottky contact changes to n-type ohmic contact.

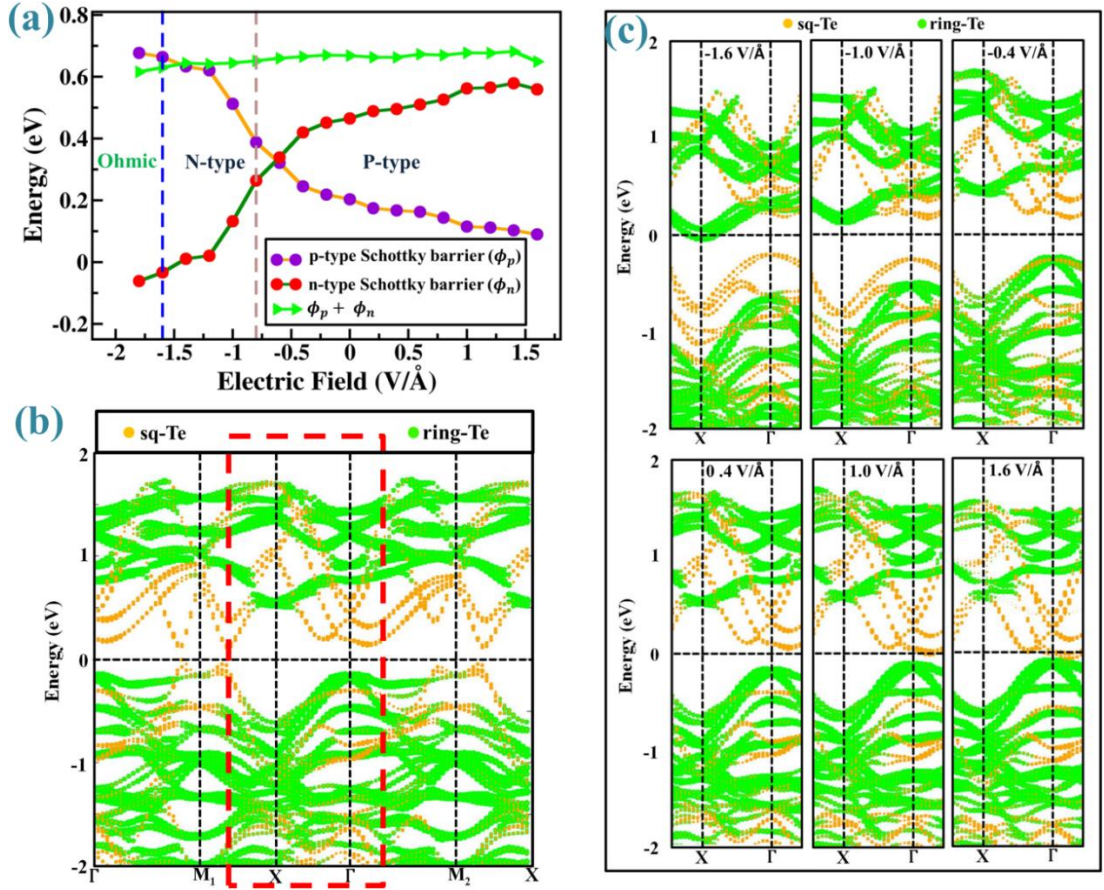

**Fig. 9(a)** The dependence of the Schottky barrier height of sq-Te/ring-Te heterostructure on the electric field. **(b)** The electronic band structure of heterostructure at zero electric field with highlighted band structure around Γ and X point. **(c)** Variation of band structures around Γ and X point with the electric field.

Next, we have plotted the projected band structure of vdWH at different electric fields (Fig. 9(b, c)). It depicts the movement of the Fermi level towards the VBM and CBM of ring-Te under the positive and negative electric fields, respectively. Such a movement of the Fermi level is attributed to the charge redistribution that occurs at the interface to attain the equilibrium when we apply the external electric field. This charge redistribution has been analyzed in terms of CDD at different electric fields (Fig. 10). The negative electric field promotes the charge transfer from sq-Te to ring-Te, leading to the movement of Fermi level towards CBM of ring-Te (n-type contact), while in the case of a positive electric field, more charge is depleted around the ring-Te, making a p-type contact at the interface of heterojunction.

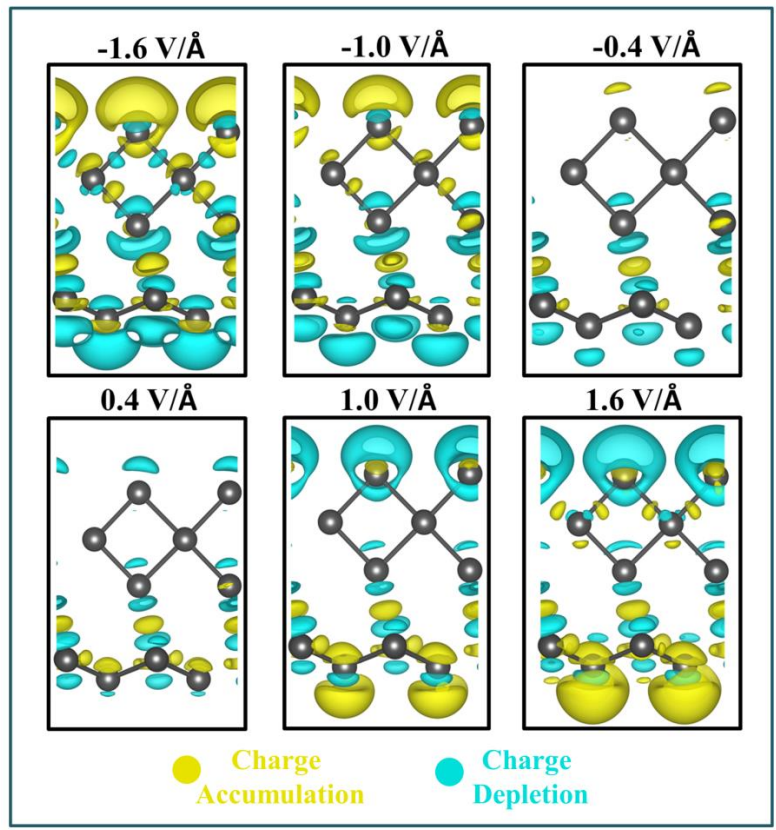

**Fig. 10** Variation of charge density difference of the sq-Te/ring-Te heterostructure with an applied electric field. The isosurface value is 0.0005 eÅ$^{-3}$.

## Conclusions

In summary, 2D novel ring structure of Te, namely, ring-Te has been reported using first principles theory. The phonon and AIMD simulations confirm its stability. It exhibits highly anisotropic mechanical properties with Young's Modulus ~3 times lower in y-direction as compared to x-direction. The ring-Te has semiconducting nature with an indirect band gap of 0.69 eV (1.16 eV) at PBE (HSE06) level of theories. The electronic structure is further tunable with applied strain that induces an indirect-to-direct band gap transition at 16% uniaxial strain in y-direction. The higher carrier mobility of holes (~$10^3 cm^2 V^{-1} s^{-1}$) than electrons (~$10^2 cm^2 V^{-1} s^{-1}$) reveals its p-type semiconducting nature. The high optical absorbance of UV-visible spectrum and low exciton binding energy (~0.35 eV) makes ring-Te as a promising material for nanodevice applications. Further, we construct vdWH (sq-Te/ring-Te) of ring-Te with topological Dirac semimetal (sq-Te). At equilibrium, the vdWH has a p-type Schottky contact with a barrier height of 0.20 eV. The transformation from p-type Schottky contact to n-type contact at -0.8 V/Å and then to ohmic contact at -1.6 V/Å with an external applied electric field reveals its applications in tunable nanodevices. Our findings could signify the exploration of 2D vdWHs composed of semiconducting materials contacted by topological semimetals for next-generation electronic device applications.

## Acknowledgments

JS is thankful to CSIR for providing the Senior Research Fellowship. The computational facility available at the Central University of Punjab, Bathinda is used to obtain the results presented in this paper. We also like to acknowledge the Mukesh Jakhar for his helpful suggestions.